\documentclass[a4paper]{article}

\usepackage{INTERSPEECH2021}

\usepackage{subfig}
\usepackage{float}
\usepackage{booktabs}

\title{Speaker-Aware Mixture of Mixtures Training for Weakly Supervised Speaker Extraction}
\name{Zifeng Zhao$^1$, Rongzhi Gu$^1$, Dongchao Yang$^1$, Jinchuan Tian$^1$, Yuexian Zou$^{1,*}$\thanks{* Corresponding author.}}
\address{$^1$ADSPLAB, School of ECE, Peking University, Shenzhen, China}
\email{\{zhaozifeng, dongchao98, tianjinchuan\}@stu.pku.edu.cn, \{1701111335, zouyx\}@pku.edu.cn}

\begin{document}

\maketitle
\begin{abstract}
  Dominant researches adopt supervised training for speaker extraction, while the scarcity of ideally clean corpus and channel mismatch problem are rarely considered. To this end, we propose speaker-aware mixture of mixtures training (SAMoM), utilizing the consistency of speaker identity among target source, enrollment utterance and target estimate to weakly supervise the training of a deep speaker extractor. In SAMoM, the input is constructed by mixing up different speaker-aware mixtures (SAMs), each contains multiple speakers with their identities known and enrollment utterances available. Informed by enrollment utterances, target speech is extracted from the input one by one, such that the estimated targets can approximate the original SAMs after a remix in accordance with the identity consistency. Moreover, using SAMoM in a semi-supervised setting with a certain amount of clean sources enables application in noisy scenarios. Extensive experiments on Libri2Mix show that the proposed method achieves promising results without access to any clean sources (11.06dB SI-SDRi)\footnote{Some audio samples of the model's output are available at our page: https://zhazhafon.github.io/demo-samom/}. With a domain adaptation, our approach even outperformed supervised framework in a cross-domain evaluation on AISHELL-1. 

\end{abstract}

\section{Introduction}
    Speech separation is a fundamental component in many speech processing systems, for example, acting as a front-end module for robust automatic speech recognition (ASR). Without such a front-end, the performances of downstream tasks may deteriorate greatly, especially when an interfering speaker exists. 
    
    Over the decades, lots of efforts have been made to crack this problem. One direction is to extract target speech with the auxiliary of an enrollment utterance from the target speaker. Following supervised learning paradigm\cite{Overview}, dominant researches formulate speaker extraction as a supervised learning problem, based on which various deep models were proposed to advance the best performance\cite{SpeakerBeam}\cite{VoiceFilter}\cite{SpEx}. In such a framework, artificially generated multi-speaker mixtures and corresponding clean sources are given as sample pairs for training. Informed by an additional enrollment utterance, a deep model consumes the input and extracts the target out of the mixture, such that the output estimate approximates target speaker's speech.
    
    Such a mix-and-separate paradigm, however, has two major drawbacks. First, corpus with adequate clean utterances is required, serving as training ground truth as well as for simulating input mixtures. Second, even if abundant simulated data is available, the model's performance in real-world scenario can still be poor, since there is usually a channel mismatch between the simulated data and the target domain.
    
    \begin{figure}[t]
      \centering
      \subfloat[During training]{
        \label{sfig:MixIT-train}
        \includegraphics[height=1.49cm]{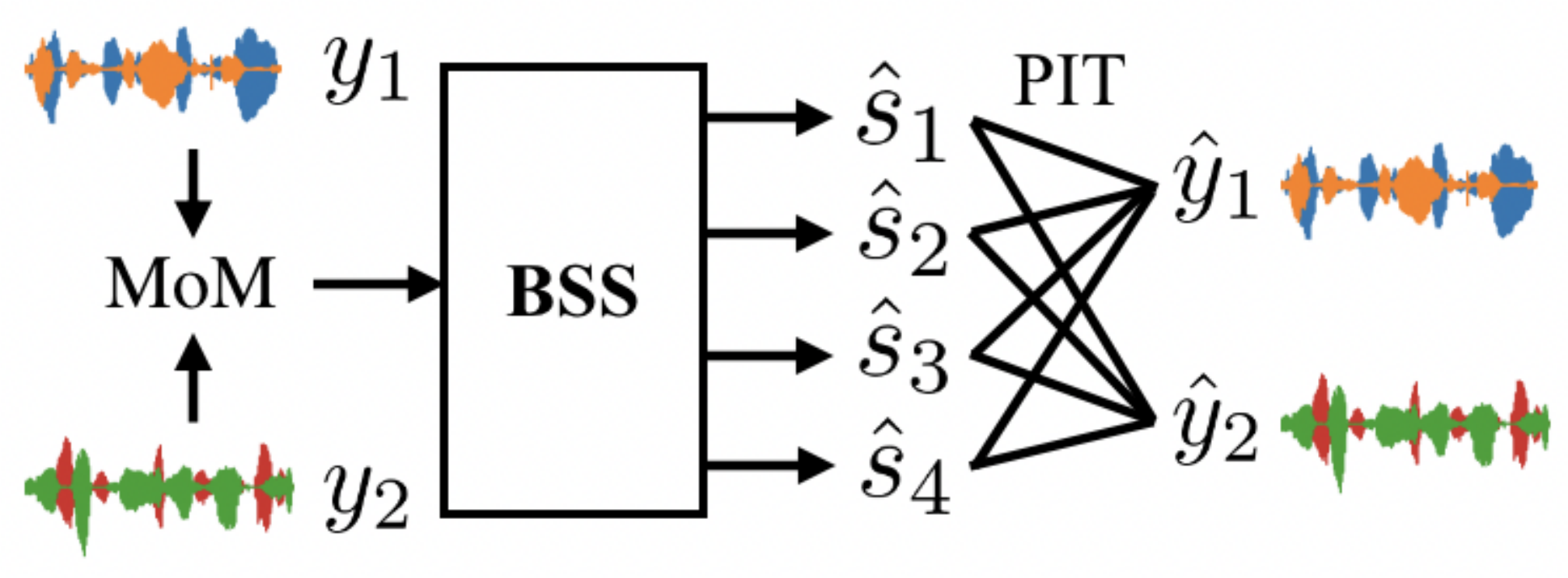}}
      \subfloat[During inference]{
        \label{sfig:MixIT-infer}
        \includegraphics[height=1.5cm]{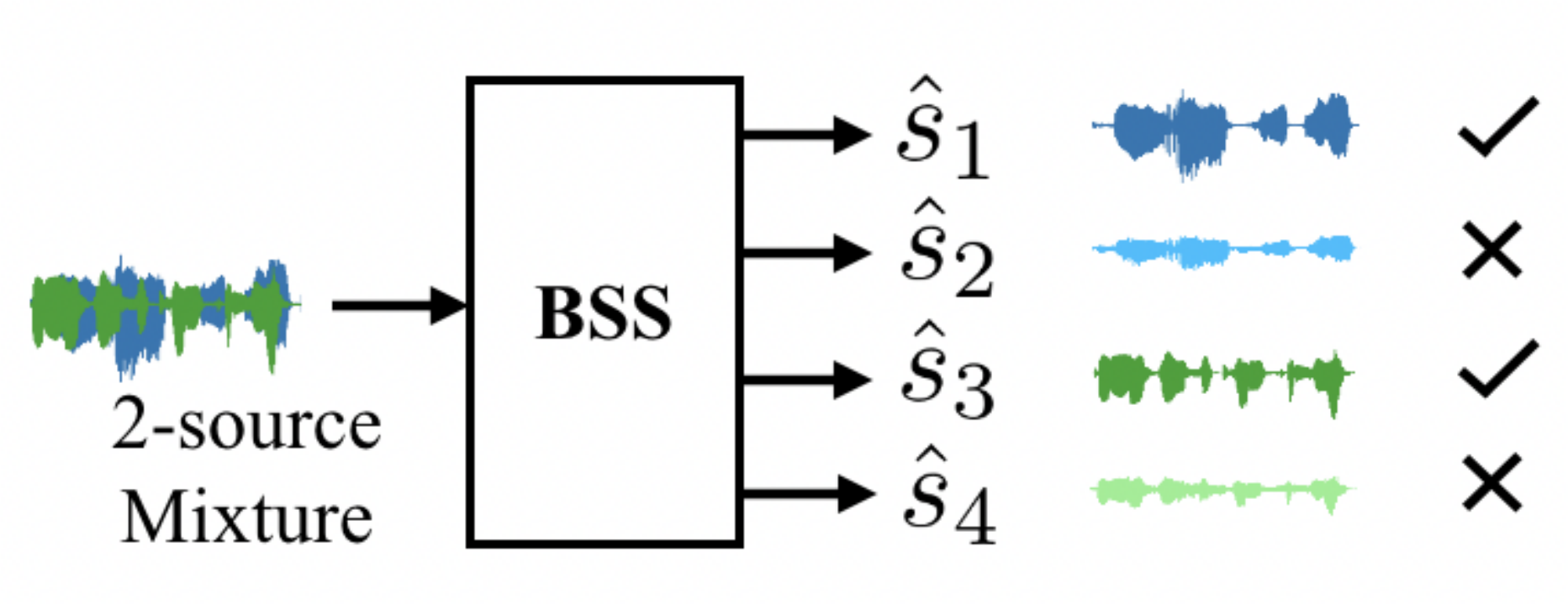}}
      \caption{MixIT for BSS models. A mismatch of the source number between (a) training and (b) inference may result in over-separaion problem.}
      \label{fig:MixIT}
    \end{figure}
    
    A feasible solution to these problems is to ameliorate current fully supervised framework with unsupervised or weakly supervised learning. Wisdom, et al. proposed mixture-invariant training (MixIT) for blind source separation (BSS)\cite{MixIT}\cite{MixIT-adapt}, where the model consumes a mixture of mixtures, and estimates multiple sources as outputs by one pass, one output channel for each latent source. In this method, a mismatch of output channel number may exist between training and inference, and hence lead to suboptimal performance due to over-separation, as illustrated in Figure 1. Very recent works also explored utilizing weak speaker labels to train guided source separation (GSS) models, especially by making use of a pretrained speaker encoder\cite{WeakSupervision}\cite{WeaklyLabelledScene} to form a speaker identity loss. An extra remix loss is also used to guarantee the mixture consistency\cite{ConsistencyConstraint}. These approaches, however, require an additional pretrained model, and may result in under-separation due to the robustness of the speaker encoder.
    
    \begin{figure*}[ht]
      \centering
      \includegraphics[width=\linewidth]{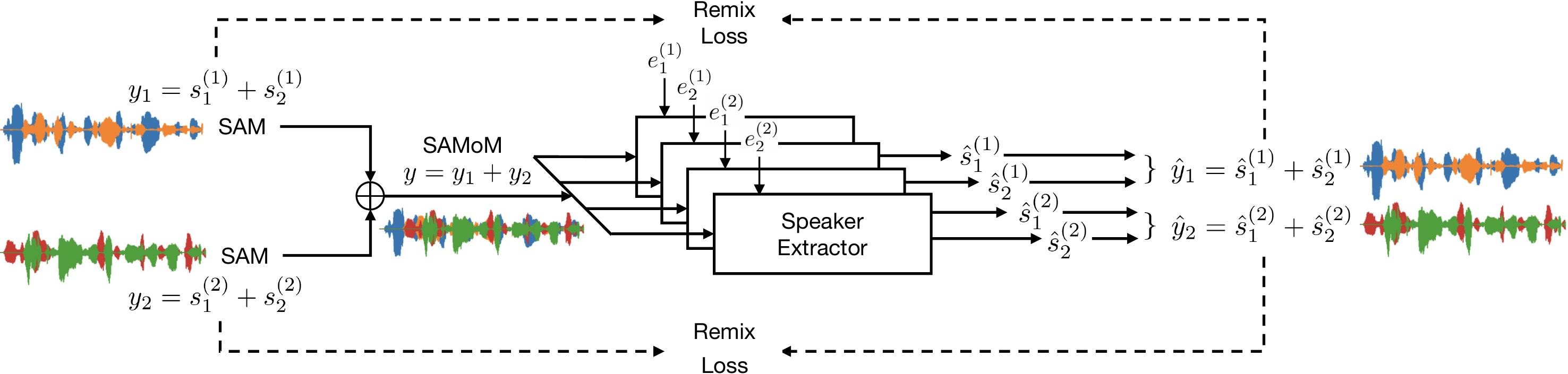}
      \caption{The proposed SAMoM training framework.}
      \label{fig:SAMoM}
    \end{figure*}
    
    In this paper, we propose speaker-aware mixture of mixtures training (SAMoM), by making advantages of the speaker identity consistency among target source, enrollment utterance and target estimate, to weakly supervise the training of a speaker extraction model. Without access to clean sources, the input is constructed by mixing up different speaker-aware mixtures (SAMs), each contains multiple speakers with their identities known and corresponding enrollment utterances available. Informed by enrollment utterances, target speech is extracted from the input one by one, such that the estimated targets can approximate the original SAMs after a remix in accordance with the identity consistency. When a certain amount of clean speech is available, SAMoM can be further extended for noisy scenarios by substituting one of the SAM with a single-speaker clean speech, while the other ingredient of the input is still a SAM, but a noisy recording with ambient sound. The input mixture is then the sum of a clean speech and a noisy SAM, while other parts of the training remains the same as the naïve SAMoM framework, forming a semi-supervised training paradigm.
    
    The rest of paper is organized as follows. Section 2 reviews speaker extraction. Section 3 presents the proposed framework and its noisy extension. Experiments and detailed results are reported in Section 4. We conclude this paper in Section 5.

\section{Target speaker extraction}
    \subsection{Problem formulation}
    In anechoic setup, the speech mixture is a linear combination of speakers' speech and ambient noise:
        \begin{equation}
            y = \sum_{j=1}^J s_j + n
            \label{equ:Mixture}
        \end{equation}
    where $J$ is the number of existing speakers, $s_j$ for $j = 1,…,J$ is the speech signal of the $j$th speaker, $n$ is the additive noise and $y$ is the speech mixture. Speaker extraction is essentially a guided speech separation, where the auxiliary information is an enrollment utterance of the target speaker:
        \begin{equation}
            \hat{s}_t = SpkExtr(y|e_t;\theta) \label{equ:SpkExtr}
        \end{equation}
    where $e_t$  is the enrollment utterance,  $SpkExtr()$  denotes a speaker extraction model with parameter $\theta$ and the output $\hat{s}_t$ is the target speech estimate. Following supervised separation \cite{Overview}, mainstream researches treat speech extraction as a supervised learning problem, in which abundant clean sources are given as ground-truth targets. Speech mixtures are artificially generated by mixing up these clean sources according to Eq. \ref{equ:Mixture}. Then mixtures and clean sources are used as sample pairs for supervised training, acting as model's inputs and labels respectively.
    
    
\section{Methods}
    
    \subsection{Mixture of mixtures}
    Supervised speech separation models are trained on mixture of sources (MoS), in which model's input is generated by mixing up different clean sources. Differently, the mixture of mixtures (MoM) paradigm is to construct an input mixture by mixing up different speech mixtures, and train the model to reconstruct those original mixtures by a properly designed loss function, such that the model acquires an ability to separate a mixture of sources for inference. Some previous researches adopted mixture of mixtures as their model's input, especially in audio-visual speech separation\cite{Co-separating} and unsupervised BSS\cite{MixIT}\cite{MixIT-adapt}\cite{TS-MixIT}, proving the effectiveness of such a paradigm. 
    
    \subsection{Speaker-aware mixture of mixtures training}
    To eliminate the over-separation problem introduced in Section 1, we make use of auxiliary speaker information and identity consistency. As illustrated in Figure \ref{fig:SAMoM}, the proposed speaker-aware mixture of mixtures training (SAMoM) framework can be divided into three phases: mixture generation for the creation of input audio, speaker extraction for target source estimation and SAM remix for remixing estimates to calculate a signal-level loss function. The overall framework is a weakly supervised learning and has no access to any clean sources.
    
	\noindent\textbf{Mixture Generation} Speaker-aware mixture (SAM) is used as a basic material for training in the proposed framework. Generally, a SAM is a mixture consisting of speech from multiple speakers, with their identities known and some enrollment utterances available, both of which are utilized as weak labels during training. The enrollment utterance provides target-related clue for speaker extraction, while the speaker identity is used to guide the subsequent remix process. Note that in naïve SAMoM, we assume no noise interference and thus the SAM is a linear combination of speech from different speakers:
	    \begin{equation}
	        y_i = \sum_{j=1}^{J_i} s^{(i)}_j
	        \label{equ:SAM}
	    \end{equation}
	where $s^{(i)}_j$ for $j=1,…,J_i$ is the speech signal from speaker $j$. $J_i$ is the speaker number in the $i$th SAM and $y_i$ denotes the $i$th SAM. With multiple SAMs available, the input is generated by:
	    \begin{equation}
	        y = \sum_{i=1}^{N} y_i
	        \label{equa:SAMoM}
	    \end{equation}
    where $N$ is the number of SAMs, $y$ is the input audio to the speaker extraction model. An example of two SAMs each with two sources ($N=2, J_1=J_2=2$) is depicted in Figure \ref{fig:SAMoM}.
    
    \noindent\textbf{Target Speaker Extraction} By informing the model of enrollment utterances that belonging to different speakers, the corresponding target speech is extracted from the input mixture one by one following Eq. \ref{equ:SpkExtr}:
        \begin{equation}
            \hat{s}^{(i)}_j = SpkExtr(y|e^{(i)}_j;\theta), \forall i, \forall j
        \end{equation}
    where $\hat{s}^{(i)}_j$ and $e^{(i)}_j$ are the speech estimate and enrollment utterance for the $j$th speaker in the $i$th SAM. Technically, this process can be done either in sequence or in parallel for different speakers, as long as the extraction for them is uncorrelated. Besides, correlated extraction methods (e.g. recursive separation\cite{RecursiveSS}) can also be employed, but we leave this to future research.
    
    \noindent\textbf{SAM Remix}  According to Sec. 2, for a certain speaker $k$, the target source estimate $\hat{s}_k$ is extracted from a SAM containing the source $s_k$, with speaker's enrollment utterance $e_k$. There is always a speaker identity consistency among these three signals: the enrollment utterance $e_k$, the target source estimate $\hat{s}_k$ and the target source $s_k$. In the last stage, estimated sources are remixed in accordance with such a consistency, so that the remixed mixtures can approximate the original SAMs:
    	\begin{equation}
    	    \hat{y}_i = \sum_{j=1}^{J_i} \hat{s}^{(i)}_j
    	\end{equation}
    where notations are consistent with Eq. \ref{equ:SAM}. Take Figure \ref{fig:SAMoM} as an example, $\hat{s}^{(1)}_1$ and $\hat{s}^{(1)}_2$ are extracted fom $y_1$ for speaker 1 and speaker 2 with their enrollment utterances $e^{(1)}_1$ and $e^{(1)}_2$. According to the identity consistency, $\hat{s}^{(1)}_1$ and $\hat{s}^{(1)}_2$ are remixed to form $\hat{y}_1$ so as to reconstruct $y_1$. Finally, the remix loss is formed by applying a negative scale-invariant signal-to-distortion ratio (SI-SDR)\cite{SISDR} between the original SAMs and the remixed SAMs:
        \begin{equation}
            L = \frac{1}{N}\sum_i^N{L_i}
        \end{equation}
        \begin{equation}
            L_i = l(y_i,\hat{y}_i)
        \end{equation}
        \begin{equation}
            l(x,\hat{x}) = -10log_{10}(\frac{||\alpha x||^2}{||\alpha x-\hat{x}||^2}), \alpha = \frac{<x,\hat{x}>}{||x||^2}
        \end{equation}
    
    \subsection{Extension to noisy scenario}
    
    \begin{figure}[h]
      \centering
      \includegraphics[width=\linewidth]{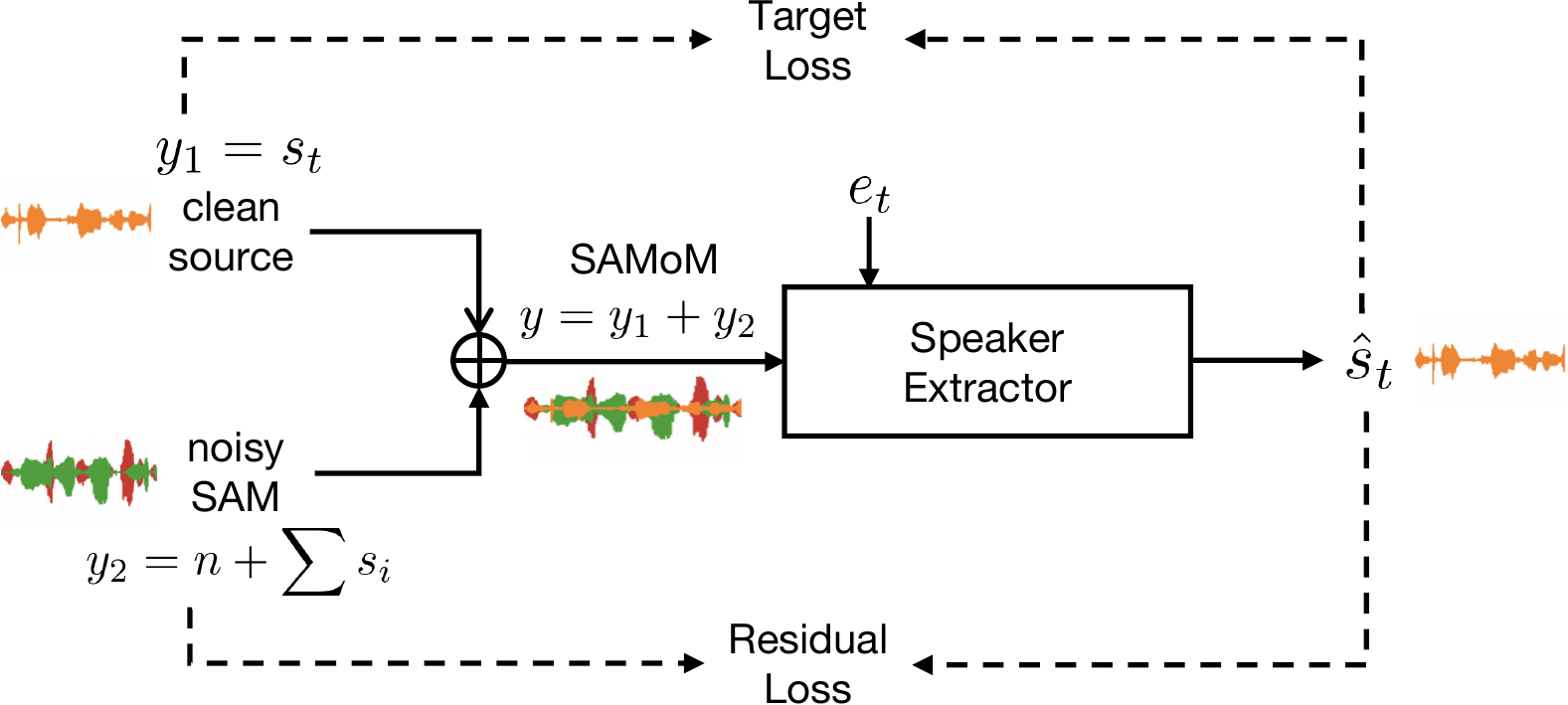}
      \caption{Extension to noisy scenario.}
      \label{fig:nSAMoM}
    \end{figure}
    
    The proposed SAMoM framework can be extended to a noisy setup for more general applications, but this may require a certain amount of clean sources, resulting in a semi-supervised training paradigm. As depicted in Figure \ref{fig:nSAMoM}, one of the SAM is substituted by a clean speech with a single speaker present, while the other ingredient of the input is still a SAM, but a noisy recording with ambient sound. The only target speaker is set to be that of the clean source and the input is the sum of a clean speech and a noisy mixture. A source estimate is extracted from the input given the enrollment utterance of the target speaker. The loss function is composed of a target loss and a residual loss:
        \begin{equation}
            L = \frac{1}{2} (L_{\textit{target}}+L_{\textit{residual}})
        \end{equation}
        \begin{equation}
            L_{\textit{target}} = l(s_t, \hat{s}_t)
        \end{equation}
        \begin{equation}
            L_{\textit{residual}} = l(y_2, y-\hat{s}_t)
        \end{equation}
    where $s_t$ is the clean source, $\hat{s}_t$ is the target source estimate, $y-\hat{s}_t$ is the estimation for the residual signal.
    
\section{Experiments}
    
    \subsection{Datasets}
    The proposed framework is evaluated under three different tasks. All audios  are downsampled to 8kHz in advance, and mixtures are truncated to 'minimum' mode.
    
	In the first task, the proposed framework is trained with only weak speaker labels and compared with fully supervised methods. We use Libri2Mix dataset\cite{LibriMix} for this task. \textit{train-100} is used as our training set, and \textit{dev} and \textit{test} subsets for validation and evaluation. Different from BSS, speaker extraction requires additional enrollment utterances. To this end, we utilize an enrollment list\footnote{https://github.com/BUTSpeechFIT/speakerbeam/tree/main/egs/libri2mix} for Libri2Mix. Note that speech mixtures with three or even more sources (e.g. Libri3Mix) are also compatible with the proposed framework. 

	The second task is a cross-domain evaluation. We created a dataset with channel characteristics different from the first task. The proposed dataset is referred to as \textit{aishell1-2mix}, which is simulated based on AISHELL-1\cite{AISHELL-1}. Test audios are directly generated by mixing up 2 randomly selected utterances of different speakers, without scaling. Note that more complicated mixing tricks can be used (e.g.random SNR sampling\cite{DPCL} or loudness control\cite{LibriMix}). Both \textit{dev} and \textit{test} subsets of AISHELL-1 are used to construct our evaluation set \textit{eval}, while the training set is not used. More details of the proposed \textit{aishell1-2mix} are listed in Table~\ref{tab:aishell1-2mix}. 
 
    \begin{table}[h]
    \centering
    \begin{tabular}{@{}lcc@{}}
    \toprule
                          & \textbf{Libri2Mix / test set} & \textbf{\textit{aishell1-2mix} / eval set} \\ \midrule
    \textbf{\#Speakers}   & 40                       & 60                           \\
    \textbf{\#Utterances} & 3000                     & 2500                         \\
    \textbf{Hours}        & 11                       & 2.08                         \\
    \textbf{Language}     & English                  & Chinese                      \\ \bottomrule
    \end{tabular}
    \caption{A comparison between the test set of Libri2Mix and the evaluation set of aishell1-2mix.}
    \label{tab:aishell1-2mix}
    \vspace{-2em} 
    \end{table}
    
    In the last task, the noisy extension of the proposed framework is evaluated. Noise from WHAM! dataset\cite{WHAM} is used together with speech from task 1 to simulate single-speaker noisy speech and two-speaker noisy speech, for semi-supervised training and evaluation, respectively.

    \subsection{Network configuration}
    The TD-SpeakerBeam\cite{TD-SpeakerBeam} is employed for our experiments, which combines the speaker clue fusion mechanism of previous works on SpeakerBeam\cite{SpeakerBeam-SA} and the time-domain convolutional separation network in Conv-TasNet\cite{Conv-TasNet}. It is chosen such that we can fairly compare our results with those by a similar BSS network (Conv-TasNet) using MixIT.
    
    The hyper-parameters are set as follows: N=512, L=16, B=128, H=512, R=3, X=8 for the time-domain convolutional separator; The auxiliary network is composed of an encoder and a single convolution block; A multiplicative adaptation layer is embedded in the 7th layer of the separator, where a 256 dimensional speaker embedding output of the auxiliary network is injected to the separator. Conv-TasNet is used for BSS MixIT baseline and the hyperparameters and network architecture are set identical with the separator above. All models are implemented using the Asteroid toolkit\cite{Asteroid} and trained for 100 epochs with Adam optimizer \cite{Adam}. The learning rate is initially set to be $10^{-3}$ and halved if validation error does not decrease in 10 consecutive epochs. During training, both input mixtures and enrollment speech are 3-second audio segments that are randomly cut from the original utterances. While for inference, full-length utterances are used.
    
    \subsection{Results}
    
    For a more complete evaluation, we compare different methods on four metrics\footnote{https://github.com/fgnt/pb\_bss}: two signal-level metrics (SI-SDRi and SDRi), one speech intelligibility metric (STOI) and one speech quality metric (PESQ).
    
\begin{table}[t]
\begin{tabular}{@{}lcccc@{}}
\toprule
                     & \textbf{SI-SDRi (dB)} & \textbf{SDRi (dB)} & \textbf{STOI} & \textbf{PESQ} \\ \midrule
\textbf{sup BSS}     & \textbf{13.40}       & \textbf{13.82}    & \textbf{0.92} & 2.74          \\
\textbf{sup SpkExtr} & 12.86                & 13.40             & 0.90          & \textbf{2.75} \\ \midrule
\textbf{unsup MixIT} & 5.72                 & 6.92              & 0.79          & 1.98          \\
\textbf{SAMoM}       & 8.97                 & 9.80              & 0.85          & 2.28          \\
\textbf{+Adaptation} & \textbf{11.06}       & \textbf{11.64}    & \textbf{0.88} & \textbf{2.41} \\ \bottomrule
\end{tabular}
\caption{Performance of different training methods for BSS and speaker extraction on Libri2Mix.}
\label{tab:task1}
\end{table}

\begin{figure}[h]
      \centering
      \subfloat[Two-speaker mixture]{
        \label{sfig:MixIT-train}
        \includegraphics[height=1.5cm]{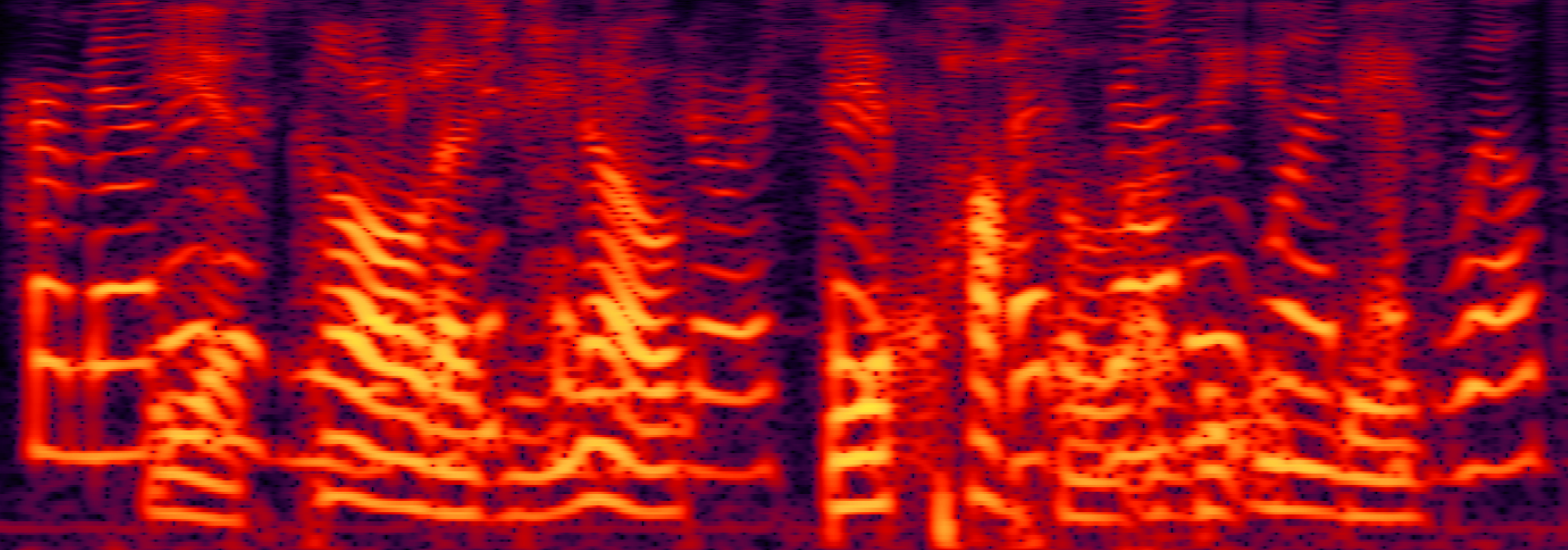}}
        
      \subfloat[Source of speaker 1]{
        \label{sfig:MixIT-infer}
        \includegraphics[height=1.35cm]{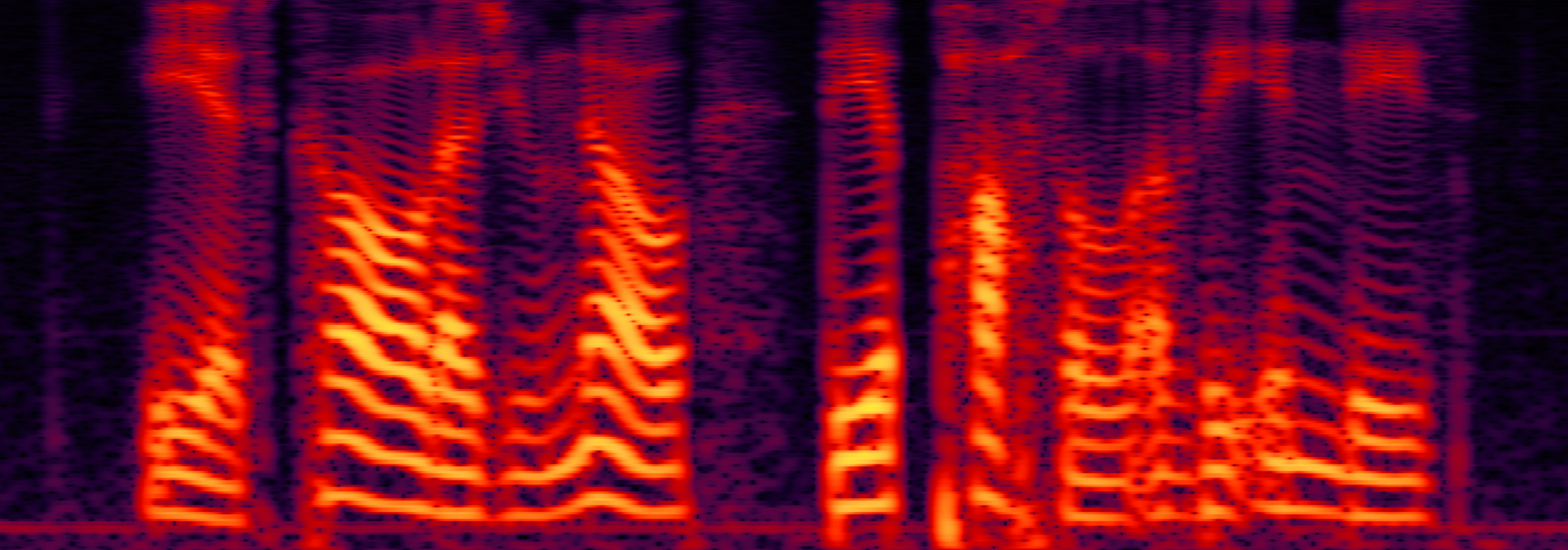}}
      \subfloat[Source of speaker 2]{
        \label{sfig:MixIT-infer}
        \includegraphics[height=1.35cm]{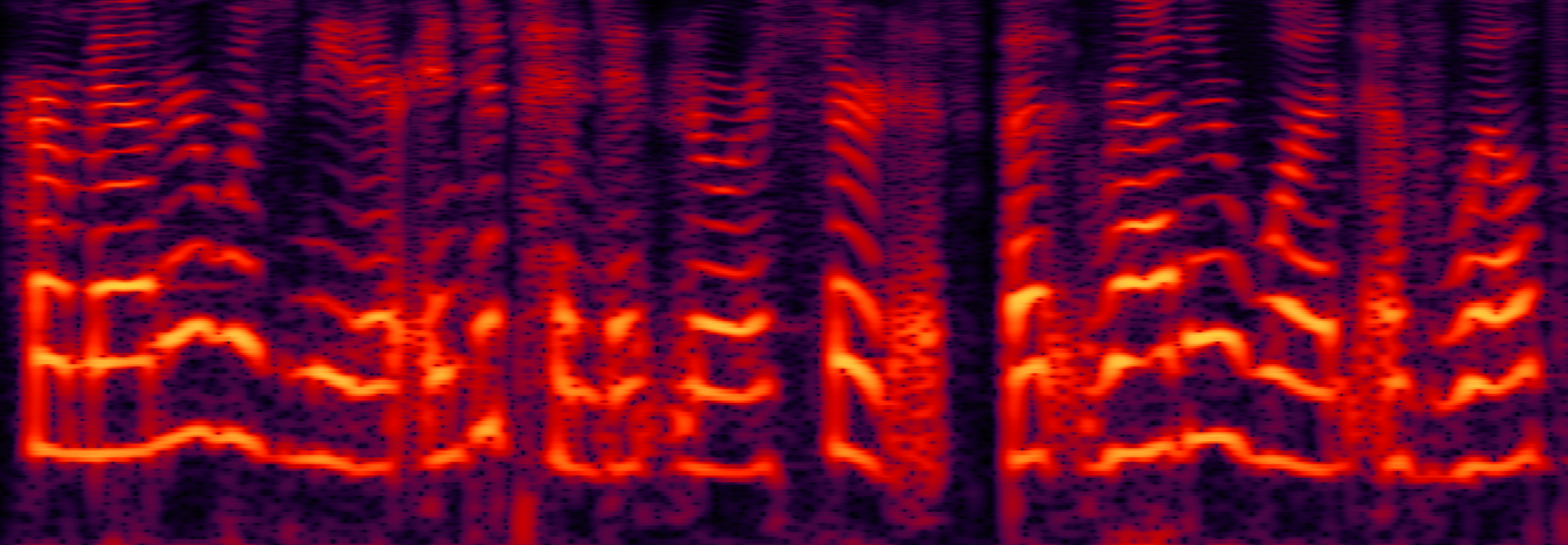}}
        
      \subfloat[Estimate of speaker 1]{
        \label{sfig:MixIT-infer}
        \includegraphics[height=1.35cm]{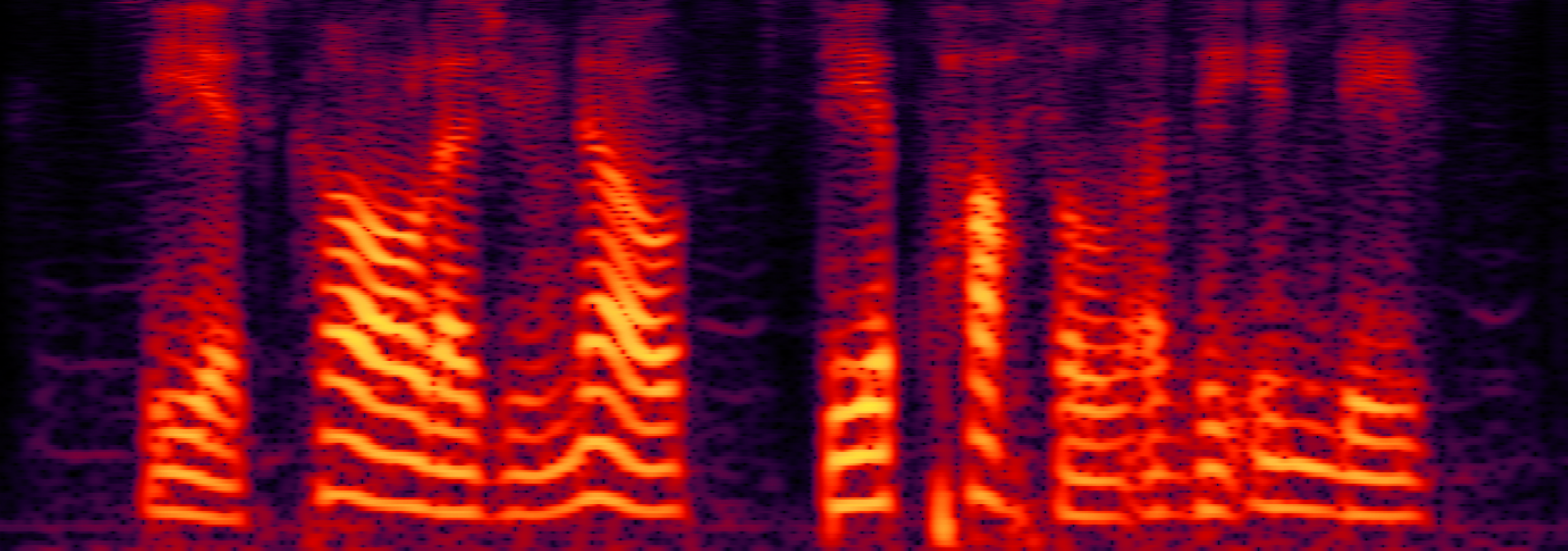}}
      \subfloat[Estimate of speaker 2]{
        \label{sfig:MixIT-infer}
        \includegraphics[height=1.35cm]{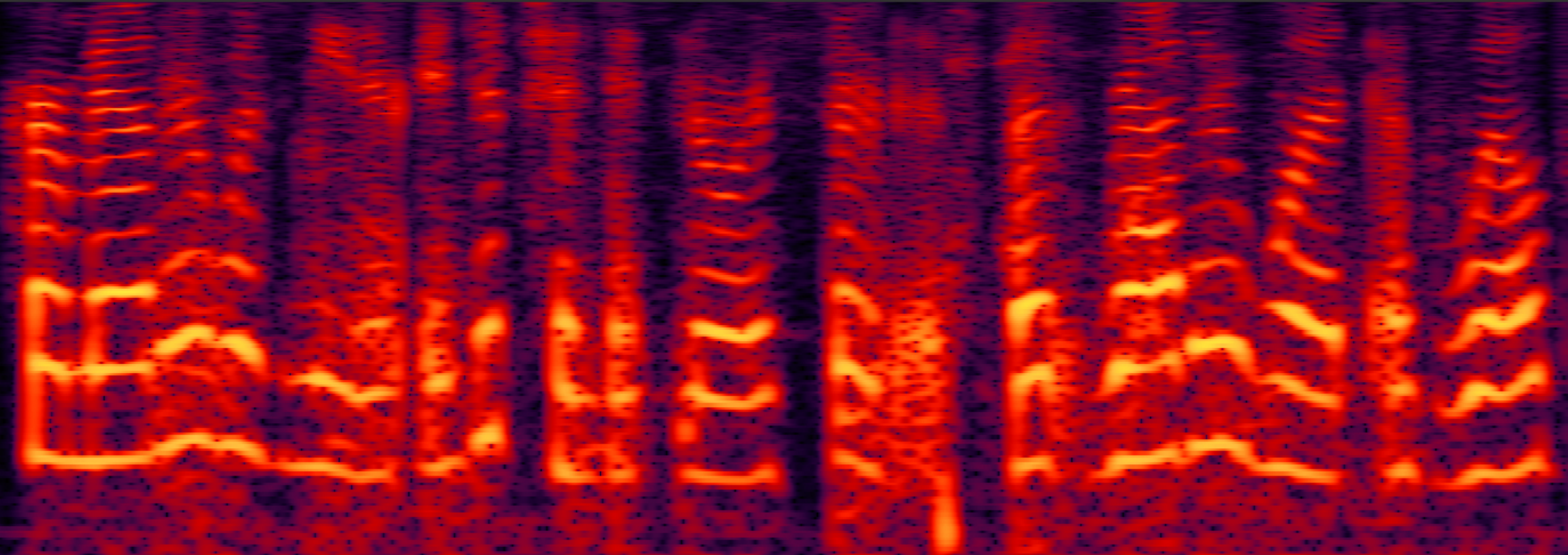}}
      \caption{Spectrograms of (a) two-speaker mixture, (b)(c) target source signals and (d)(e) estimated signals on Libri2Mix.}
      \label{fig:spectrogram}
    \vspace{-2em}
    \end{figure}

    \noindent\textbf{Proposed method and baselines} In the first task, we compared our method with several baselines on Libri2Mix: (1) \textit{sup BSS}: BSS with supervised training, (2) \textit{sup SpkExtr}: speaker extraction with supervised training and (3) \textit{unsup MixIT}: BSS with MixIT unsupervised training. Permutation-invariant training\cite{PIT}\cite{uPIT} is adopted for BSS models (\textit{sup BSS} and \textit{unsup MixIT)} during both training and inference. While for speaker extraction models, the enrollment list introduced in Section 4.1 is used. Results are reported in Table \ref{tab:task1}. As illustrated in the first two rows, \textit{sup SpkExtr} is slightly inferior to \textit{sup BSS}, probably due to that speaker extraction comes across with speaker bias in some enrollment utterances; MixIT\cite{MixIT} with purely unsupervised training (\textit{unsup MixIT}) achieved a SI-SDRi of 5.72dB, which is far less than the fully supervised BSS baseline (13.40dB SI-SDRi). This is because a mismatch of output channel number exists between training and inference, which leads to over-separation during testing, as depicted in Figure \ref{fig:MixIT}. Our proposed method (\textit{SAMoM}) significantly outperforms \textit{unsup MixIT} by more than 3dB in terms of SI-SDRi. Furthermore, since \textit{SAMoM} does not require any clean sources for training, it can adapt to the testing data through an additional fine-tuning. This is done by training the model with weakly-supervised learning on the test set for 20 more epochs. The learning rate is set to be 0.0001 and halved if validation error does not decrease in 2 consecutive epochs. With such an additional adaptation on the test set, \textit{SAMoM+Adaptation} achieved a SI-SDRi of 11.06 dB, which is close to that of \textit{sup SpkExtr} (12.86dB SI-SDRi). A sample data of evaluation results from \textit{SAMoM+Adaptation} is depicted in Figure \ref{fig:spectrogram}.

\begin{table}[h]
\begin{tabular}{@{}lcccc@{}}
\toprule
\textbf{}            & \textbf{SI-SDRi (dB)} & \textbf{SDRi (dB)} & \textbf{STOI} & \textbf{PESQ} \\ \midrule
\textbf{Sup-init}    & 1.99                 & 2.65              & 0.68          & 1.77          \\
\textbf{+Adaptation} & 4.56                 & 5.48              & 0.73          & 2.06          \\
\textbf{SAMoM-init}  & 0.73                 & 1.97              & 0.66          & 1.72          \\
\textbf{+Adaptation} & \textbf{5.86}        & \textbf{6.64}     & \textbf{0.75} & \textbf{2.12} \\ \bottomrule
\end{tabular}
\caption{Cross-domain evaluation on aishell1-2mix.}
\label{tab:task2}
\end{table}
\vspace{-2em} 

    \noindent\textbf{Cross domain evaluation} The second task is to show model's generalization ability when applied to a new scenario with completely different channel characteristics. Two base models were used in this task: (1) \textit{Sup-init}: fully supervised training, equivalent to \textit{sup SpkExtr} in task 1, (2) \textit{SAMoM-init}: weakly supervised training, equavalent to \textit{SAMoM} in task 1. Both of the models were trained on Libri2Mix with the same setups as task 1, and evaluated on the proposed \textit{aishell1-2mix}. As illustrated in Table \ref{tab:task2}, although \textit{Sup-init} performs better than \textit{SAMoM-init}, both of their performance are very poor when confronted with a different channel characteristic. To ease such a data mismatch, a domain adaptation can be done by fine-tuning base models in the target domain through SAMoM training (\textit{+Adaptation}). To this end, base models are fine-tuned on \textit{aishell1-2mix} for 20 epochs, with an initial learning rate of 0.001 and halved at the antepenultimate epoch. With fine-tuning, the performance of \textit{Sup-init} and \textit{SAMoM-init} increased to 4.56dB and 5.86dB SI-SDRi, respectively, showing that such a domain adaptation can play a crucial role for cross-domain inference. 

\begin{table}[]
\begin{tabular}{@{}lcccc@{}}
\toprule
\textbf{}           & \textbf{SI-SDRi (dB)} & \textbf{SDRi (dB)} & \textbf{STOI} & \textbf{PESQ} \\ \midrule
\textbf{Supervised} & \textbf{10.79}       & \textbf{11.51}    & \textbf{0.83} & \textbf{2.15} \\
\textbf{Proposed}   & 9.55                 & 10.26             & 0.81          & 1.99          \\ \bottomrule
\end{tabular}
\caption{Performance under noisy condition.}
\label{tab:task3}
\end{table}

    \noindent\textbf{Noisy extension} As illustrated in Table \ref{tab:task3}, SAMoM's semi-supervised extension achieved a SI-SDRi of 9.55dB, which is close to the fully supervised model (10.79dB). This suggests the effectiveness of the proposed method. 

\section{Conclusions}
    In this paper, we propose speaker-aware mixture of mixtures training (SAMoM), a weakly supervised learning framework for speaker extraction. The proposed method achieves considerable results with only weak speaker labels accessible. Since no clean sources are required for training, it can realize domain adaptation to reduce performance attenuation caused by channel mismatch. In addition, we extend it for noisy condition with semi-supervised learning. Extensive experiments on LibriMix and AISHELL-1 validate the effectiveness of our methods. 
    
\section{Acknowledgements}

This paper was partially supported by the Shenzhen Science \& Technology Fundamental Research Programs (No:JSGG20191129105421211).


\bibliographystyle{IEEEtran}


\end{document}